# SEMI-ANALYTICAL PERTURBATIVE APPROACHES TO THIRD BODY RESONANT TRAJECTORIES

**Joan Pau Sánchez Cuartielles**
Space Research Centre, Cranfield University, Bedfordshire, UK.
jp.sanchez@cranfield.ac.uk

**Camilla Colombo**
Aeronautics, Astronautics and Computational Engineering Unit, University of Southampton, Southampton, UK.
c.colombo@soton.ac.uk

**Elisa Maria Alessi**
Istituto di Fisica Applicata "Nello Carrara" - Consiglio Nazionale delle Ricerche (IFAC-CNR), Sesto Fiorentino (FI), Italy. em.alessi@ifac.cnr.it

In the framework of multi-body dynamics, successive encounters with a third body, even if well outside of its sphere of influence, can noticeably alter the trajectory of a spacecraft. Examples of these effects have already been exploited by past missions such as SMART-1, as well as are proposed to benefit future missions to Jupiter, Saturn or Neptune, and disposal strategies from Earth's High Eccentric or Libration Point Orbits. This paper revises three totally different descriptions of the effects of the third body gravitational perturbation. These are the averaged dynamics of the classical third body perturbing function, the Öpik's close encounter theory and the Keplerian map approach. The first two techniques have respectively been applied to the cases of a spacecraft either always remaining very far or occasionally experiencing extremely close approaches to the third body. However, the paper also seeks solutions for trajectories that undergo one or more close approaches at distances in the order of the sphere of influence of the third body. The paper attempts to gain insight into the accuracy of these different perturbative techniques into each of these scenarios, as compared with the motion in the Circular Restricted Three Body Problem.

## INTRODUCTION

In the past years, the idea of taking advantage of a sequence of resonant encounters with a given planet, or moon, in order to obtain a change on the trajectory that would otherwise require propellant consumption, has been widely considered in order to reduce the cost of interplanetary transfers. In the framework of multi-body dynamics, successive encounters with a third body, even if well outside of its sphere of influence, noticeably alter the trajectory of a spacecraft. A successful example of this is represented by the SMART-1 lunar mission, which used the perturbation of the Moon to raise the perigee of its orbit sufficiently to get ballistically captured by it. Other mission concepts have been hypothesised on the same basis, such as tours of planetary moons in Jupiter, Saturn or Neptune, as well as disposal strategies from Earth's High Eccentric or Libration Point Orbits.

In this framework, the paper revises classical general perturbation approaches based on averaged dynamics developed, for example, by Brouwer [1] and Kozai [2], as well as Öpik's close encounter theory [3]. These methods have generally been applied to two extreme cases; the first is that of a spacecraft whose orbit always remains very far from the perturbing body (i.e., the ratio between the spacecraft and the third-body distance from the central planet is much smaller than 1), while the second one is that of a spacecraft that at some point along its trajectory undergoes an extremely close approach with the third body, well within its sphere of influence. However, we also seek solutions that do not fall in any of the previous situations. These are trajectories that undergo close approaches at distances in the order of the sphere of influence of the third body. We thus explore the accuracy of the classical averaging methods for this kind of scenario, but also explore a semi-analytical approach that is based instead on a linear expansion in the mass parameter $\mu$ of the Circular Restricted Three Body Problem (CR3BP) as an extension of the so-called Keplerian map methods [4].

Hence, the paper attempts to gain insight into the accuracy of the different approximations, as compared with the motion in the CR3BP. This responds to current needs in astronautics for ever more complex trajectory planning that benefit from the subtleties of multi-body dynamics (e.g., [5, 6]). An initial effort in this direction was presented in Alessi et al [7], however this conference paper goes further into the exploration of the domains of applications of each of the aforementioned perturbative techniques for different systems (i.e., Earth-Moon, Sun-Earth, Jupiter-Callisto) and orbits.

# THIRD BODY PERTURBATIONS

This section will first review the fundamental characteristics of these three very different approaches to model the influence of a third body gravitational perturbation into the Keplerian elements of a massless particle. In the following we refer as a massless particle the object for which we want to analyse the evolution of the orbit under the influence of a perturbation. This object may generally be a spacecraft, but as later described, it could also be an asteroid or comet. Clearly, the term massless makes a reference to the assumption that the mass of this object is negligible, and thus its influence into the motion of planetary bodies is ignored.

These three approaches to the third body perturbation are: third body averaged dynamics, Öpik's close encounter theory and Keplerian map approach. Note that here we refer to third body averaged dynamics as the application of averaging techniques into the *classical* third body perturbing function (i.e. [1, 2, 8]), however the averaged technique is independent of the perturbing functions.

## Single and Double Averaging Methods

A general approach for taking into account the effect of third body perturbation due to the gravitational attraction of an external body is the two body problem perturbed approach, where the perturbing potential of a third body is added onto the unperturbed two body dynamics by means of Lagrange Planetary equations. When the long-term evolution of the orbit is of interest, semi-analytical techniques, based on averaging, are used to account for the long-term and secular effects of the perturbing planet onto the two-body motion. Cook's formulation gives the secular and long-periodic effects due to luni-solar perturbation obtained through averaging over one orbit revolution of the satellite [8]. It assumes circular orbit for the disturbing bodies and considers only first terms of $a/a_D$, where $a$ and $a_D$ are respectively the spacecraft and the disturbing body semi-major axis. However, they do take into account the obliquity of the Sun and the Moon over the equator and the precession of the Moon plane due to the Earth's oblateness (in a period of 18.6 years with respect to the ecliptic). Alternatively, Chao gives another form of the averaged equations obtained from an expansion of the disturbing function from third-body perturbations up to the second order of the disturbing potential [9]. In a recent work on the study of Highly Elliptical Orbits we instead chose the approach proposed by Kauffman and Dasenbrock [10, 11], as it allows treating orbits with high eccentricity; moreover, we extended the computation of the derivatives of the disturbing function to order $12^{th}$. In the following, a brief description of the single and double averaged approach is provided, a more extended explanation can be found in [11].

The disturbing potential due to the third body perturbation is

$$R(r,r') = \frac{\mu'}{r'}\left(\left(1 - 2\frac{r}{r'}\cos\psi + \left(\frac{r}{r'}\right)^2\right)^{-1/2} - \frac{r}{r'}\cos\psi\right) \quad (1)$$

where $\mu'$ is the gravitational coefficient of the third body, $\mathbf{r}$ and $\mathbf{r'}$ are the position vectors of the satellite and the third body with respect to the central planet and $\psi$ is the angle between $\mathbf{r}$ and $\mathbf{r'}$. Kauffman and Dasenbrock [10] express the disturbing potential as function of the spacecraft's and third body's orbital elements, choosing as angular variable the eccentric anomaly $E$, the ratio between the orbit semi-major axis and the distance to the third body $r'$ on its circular orbit $\delta = a/r'$ and the orientation of the orbit eccentricity vector with respect to the third body. Under the assumption that the parameter $\delta$ is small (i.e., the spacecraft is far enough from the perturbing body), Eq.(1) can be rewritten as a Taylor series in $\delta$. The average operation in mean anomaly can then be performed, assuming that the orbital elements of the spacecraft $a$, $e$, $i$, $\Omega$ and $\omega$ are constant over one orbit revolution, to obtain:

$$\bar{R}(r,r') = \frac{\mu'}{r'}\sum_{k=2}^{\infty}\delta^k \bar{F}_k(A,B,e) \quad (2)$$

Note that many derivations of this method consider only up to term 2 of this expansion, thus limiting the use of the resulting expression to orbits with small semi-major axis with respect to the distance to the third body. Eq.(2) can be now inserted into the Lagrange equations by computing the derivatives with respect to the orbital elements, considering that $A$ and $B$ depends on $i$, $\Omega$ and $\omega$ and $\Omega', i', \omega' + f'$ which are respectively the inclination, anomaly of the ascending node and true anomaly of the third body as seen from the primary body. The derivatives are reported in [10].

As shown by Colombo [11], under the further assumption that the orbital elements do not change significantly during a full revolution of the perturbing body around the central body (i.e., Earth), the variation of the orbit over time can be approximately described through the disturbing potential double averaged over one orbit evolution of the spacecraft and over one orbital revolution of the perturbing body (either the Moon or the Sun) around the Earth:

$$\bar{\bar{R}} = \frac{\mu'}{r'}\sum_{k=2}^{\infty}\delta^k \bar{\bar{F}}_k(e,i,\Delta\Omega,\omega,i') \quad (3)$$

where $\Delta\Omega = \Omega - \Omega'$. Performing the double-averaged potential with respect to the Keplerian elements described in the Earth's centred equatorial reference system gives a more complex expression with respect to the one proposed by El'yasberg [12] and Costa and Prado [13], but it has the advantage of avoiding the simplification that Moon and Sun orbit on the same plane that would include errors in the case of a spacecraft orbiting the Earth, under the third body effects of the Sun and the Moon. For example, in the approach proposed in [11] it is possible to consider the actual ephemerides of the Moon's plane that processes due to the effect of the Earth's oblateness with a period of 18.6 years.

In this work, in order to have a common framework to assess and compare the different representation of the third body effect, we will validate the double averaged approach developed in [11] with the CR3BP. By expressing the double averaged potential with respect to the third body plane, with its *x*-axis in the direction of the perturbing body on its orbit, the doubly averaged potential loses the dependence on the anomaly of the ascending node.

Keplerian Map

What follows is a brief description of a semi-analytical model that approximates the CR3BP by means of perturbation techniques. Building on previous authors' terminology (e.g., [14-16]), the approximation presented here is referred as Keplerian Map. The key feature of this methodology, as well as in those from previous authors [14-16], is the consideration that the mass of the third body can act as a small parameter for the perturbative approach. Note that in the previous approach the small parameter was the ratio $a/r'$. Hence, the Hamiltonian of the CR3BP can be approximated, after some algebraic manipulation, as:

$$H_{iner} = K - R + O(\mu^2) \qquad (4)$$

where $\mu = \dfrac{\mu'}{\mu' + \mu_1}$ and $\mu_1$ is the gravitational coefficient of the centre body, while *K* represents the Keplerian term, namely,

$$K = \frac{1}{2}\left(p_x^2 + p_y^2 + p_z^2\right) - \frac{1}{r_b} \qquad (5)$$

with $p_x$, $p_y$ and $p_z$ as generalized momenta of the massless particle and $r_b$ the distance from the system barycentre to the massless particle. *R* represents the perturbing function, as:

$$R(r_b) = -\mu\left(\frac{1}{r_b} + \frac{\cos\theta}{r_b^2} - \frac{1}{\sqrt{1 + r_b^2 - 2r_b\cos\theta}}\right) \qquad (6)$$

where $\theta$ is the angle between the direction joining the centre body and the third body and the radius vector $\mathbf{r}_b$ measured from the system barycentre. For a comprehensive description of the derivation of these equations refer to Alessi and Sanchez [4]. Note that in order to transform the Hamiltonian of the CR3BP into the approximation described by Eqs. (4-6), the distance of the massless particle to the centre body *r* and to the perturbing third body $r_{3B}$ are approximated by means of binomial series as functions of $r_b$, $\theta$ and $\mu$ and only the linear terms in the mass parameter $\mu$ are retained. Thus, the main difference from the previous approach is that in this case no hypothesis is made on the distance to the third body, or, using the above description, $r_b$ must not necessarily satisfy $r_b \ll 1$. Future work will be devoted to demonstrate that the disturbing potential in Eq.(1) can be expressed as in Eq.(6) by changing the notation from a planet-centred dynamics into a barycentre-centred dynamics, proper of the CR3BP.

A set of Keplerian orbital elements can then be associated to the motion of the massless particle around the centre body, while Lagrange planetary equations can be used to estimate the rate of change of the latter due to the perturbation of the third body. By means of a Picard's first iteration, an estimate the changes over a fixed timespan can be estimated. Clearly, this timespan needs to be short enough so that Picard's first iteration provides a good approximation. The latter, again, puts some constraints to the applicability of this method. The complete set of equations are omitted here, but can be found in a recently published paper by the authors [4].

The Öpik's Approach

The third approach considered in this work is that of Öpik's theory [3], whose physical assumptions are the same as those used in the patched conics approach. Considering as example a spacecraft orbiting the Sun, a close encounter with a planet is described as a planetocentric hyperbolic passage, resulting in an exchange of energy and angular momentum between the satellite and the planet. The planetocentric velocity vector remains constant in modulus, but it is rotated instantaneously by an amount which depends on its magnitude, the distance of the closest approach and the mass of the planet. Because of this, the heliocentric velocity vector of the small body changes both in modulus and direction. What follows is based on the works by Valsecchi et al. (2003), Valsecchi (2006) and Valsecchi et al. (2015) [17-19].

To describe the encounter, the theory starts from the model of the CR3BP, but further simplifications are introduced. In particular, far from the planet the small body is assumed to move on an unperturbed heliocentric Keplerian orbit, whereas the encounter with the planet is modelled as an instantaneous transition from the incoming asymptote of the planetocentric hyperbola to the outgoing one. The transition takes place when the small body crosses the so-called *b*-plane, which is defined as the plane perpendicular to the incoming asymptote and centred on the planet.

To clarify the notation, in Öpik's theory the reference frame adopted is centred on the third body (i.e., the planet and usually referred as secondary in the usual CR3BP terminology), with the central body (i.e., the Sun) on the negative *X*-axis. The *Y*-axis is aligned with the velocity of the third body and the *Z*-axis is aligned with the angular momentum vector of the third body in its orbit around the centre body. As mentioned, the main assumption is that the massless particle crosses the orbit of the third body, and that the encounter is instantaneous: the reference system introduced is defined at the time of the encounter. Moreover, the hyperbolic excess velocity is denoted by **U**, and its direction is specified by two angles, $\theta$ and $\phi$, which are the angle from the heliocentric velocity of the planet to **U**, and the angle between the *Y-Z* plane and the plane containing the *Y*-axis and **U**, respectively. Note that angle $\theta$ refers here to a different angle than in the previous section. The units are taken as in the CR3BP, that is, the unit of length is the radius of the circular orbit of the third body around the centre mass, the unit of time is such that one orbital period of the third body is equal to $2\pi$ and the unit of mass is the mass of the centre body. In this definition, the mass of the third body is considered negligible and the Universal Gravitational Constant turns out to be *G*=1. Moreover, the modulus of the semi-major axis of the hyperbola is $c = m/U^2$ where *m* is the mass of the third body in units of mass of the centre body, and *U* is computed in units of speed of the third body. Using the CR3BP convention, it turns out that $m = \mu/(1-\mu)$.

Given a set of orbital elements for the massless particle, defined at a given time $t^*$, say $(a,e,i,\Omega,\omega,v^*)$, with respect to the central body in the orbital plane of the perturbing third body, the variables $U$, $\theta$ and $\phi$ are given by

$$U = \sqrt{3 - \frac{1}{a} - 2\sqrt{a(1-e^2)}\cos i}$$

$$\cos\theta = \frac{1 - U^2 - \frac{1}{a}}{2U} \quad \sin\theta = \frac{\sqrt{2 - \frac{1}{a} - a(1-e^2)\cos^2 i}}{U} \quad (7)$$

$$\sin\phi = \frac{\sin v^*}{|\sin v^*|} \frac{\sqrt{2 - \frac{1}{a} - a(1-e^2)}}{U\sin\theta}$$

$$\cos\phi = \frac{\cos(\omega+v^*)}{|\cos(\omega+v^*)|} \frac{\sqrt{a(1-e^2)}\sin i}{U\sin\theta}$$

The procedure described in what follows can be considered reliable, provided that some hypotheses are verified. In particular,

- The orbit of the massless particle about the third body at encounter must be hyperbolic, that is,

$$\mathcal{T} = \frac{1}{a} + 2\sqrt{a(1-e^2)}\cos i < 3$$

- The third body and massless particle must not be co-moving.
- The encounter must be close.
- The mass of the third body must be small. The encounter velocity *U* is in units of the velocity of the third body about the centre body, provided that the velocity of the third body is computed as if its mass were 0, introducing an error of order *m*.

We remark that our aim here is to define a numerical boundary of the last two assumptions, which are related to the fact that the encounter variables are linearized not only in the impact parameter *b*, but also in *m*.

*Practical Procedure*

Given a set of orbital elements of the massless particle with respect to the centre body, let us assume to know the time of closest approach, say $t_b$, between the massless particle and the third body in one orbital period of the former, and the corresponding relative position coordinates, say $(X_b, Y_b, Z_b)$, in the Öpik's reference frame. They can be computed by applying the procedure explained in Valsecchi et al. (2015) [19], or by propagating the classical Kepler's equation for both bodies[*] looking for the time when the relative distance attains a minimum value, and then applying a rotation around the *Z*-axis of an angle equal to $t_b$.

In the *b*-plane, the coordinates of the small body are

$$\xi = X_b \cos\phi - Z_b \sin\phi$$
$$\zeta = (X_b \sin\phi + Z_b \cos\phi)\cos\theta - Y_b \sin\theta \quad (8)$$

where $\theta$, $\phi$ are computed from Eq.(7) with $v^* = v_b$.

The motion resulting from the rotation from the incoming to the outgoing asymptote at time $t_b$ is given by

$$U' = U$$

$$\cos\theta' = \frac{(b^2 - c^2)\cos\theta + 2c\zeta\sin\theta}{b^2 + c^2}$$

$$\sin\theta' = \frac{\sqrt{[(b^2 - c^2)\sin\theta - 2c\zeta\cos\theta]^2 + 4c^2\xi^2}}{b^2 + c^2}$$

$$\cos\phi' = \frac{[(b^2 - c^2)\sin\theta - 2c\zeta\cos\theta]\cos\phi + 2c\xi\sin\phi}{(b^2 + c^2)\sin\theta'}$$

$$\sin\phi' = \frac{[(b^2 - c^2)\sin\theta - 2c\zeta\cos\theta]\sin\phi - 2c\xi\cos\phi}{(b^2 + c^2)\sin\theta'}$$

$$\xi' = \frac{\xi\sin\theta}{\sin\theta'}$$

$$\zeta' = \frac{(b^2 - c^2)\zeta\sin\theta - 2b^2 c\cos\theta}{(b^2 + c^2)\sin\theta'}$$

where $b^2 = \xi^2 + \zeta^2$.

---

[*]Recall that the orbital elements of the secondary with respect to the primary in the units adopted are (1,0,0,0,0,*t*).

The orbital elements of the new heliocentric orbit (denoted with a prime as well) are thus given by

$$a' = \frac{1}{1 - U'^2 - 2U'\cos\theta'}$$

$$e' = U'\sqrt{(U' + 2\cos\theta')^2 + \sin^2\theta'\sin^2\phi'(1 - U'^2 - 2U'\cos\theta')}$$

$$\cos i' = \frac{1 + U'\cos\theta'}{\sqrt{1 + 2U'\cos\theta' + U'^2(1 - \sin^2\theta'\sin^2\phi')}}$$

$$\cos v' = \frac{a'(1 - e'^2) - (1 + X'_b)}{e'(1 + X'_b)}$$

$$\sin v' = \frac{\sin\phi'}{|\sin\phi'|}\sqrt{1 - \cos^2 v'}$$

$$\sin(\omega' + v') = \frac{(1 + e'\cos v')Z'_b}{a'(1 - e'^2)\sin i'}$$

$$\cos(\omega' + v') = \frac{\cos\phi'}{|\cos\phi'|}\sqrt{1 - \sin^2(\omega' + v')}$$

$$\Omega' = t_b - 2\arctan\left[\frac{\sin(\omega' + v')\cos i'}{1 + \cos(\omega' + v')}\right] + \frac{(1 + e'\cos v')Y'_b}{a'(1 - e'^2)}$$

where

$$X'_b = \zeta'\cos\theta'\sin\phi' + \xi'\cos\phi'$$

$$Y'_b = -\zeta'\sin\theta'$$

$$Z'_b = \zeta'\cos\theta'\cos\phi' - \xi'\sin\phi'$$

FRAMEWORK OF APPLICATIONS: CASES

This section will demonstrate the applicability of the three methodologies previously described in three different examples. Along with the demonstration of their application, some more insights of the limits of each approach will be given.

XMM-Newton like spacecraft

We will first demonstrate the use of the double averaging method of the third body perturbation to estimate the orbital evolution for XMM spacecraft. As analysed in [11] the main effect for this orbit is due to the third body effects of the Moon and the Sun; although here only the Moon will be considered to analyse the effect of the third body. The orbital elements of the spacecraft in an Earth centred system with the *x-y* plane corresponding to the Earth-Moon plane are contained in Table 1. The averaged approach to the third body perturbation presented earlier will be compared to the propagation of the dynamic by using the CR3BP. Note that, with respect to [11] here the effect of the Earth' oblateness on the ephemerides of the Moon are neglected and the Moon is considered on a circular orbit with radius of 384,400 km, inclination of 25.7 deg with respect to the equator and ascending node of 11.5 deg. Figure 1 shows the long term evolution of the orbital elements of XMM spacecraft under the perturbing effect of the Moon using the semi-analytical double averaged method (red line) and the CR3BP (blue line). The semi-analytical techniques capture the secular and long-term variation of the orbital elements, filtering the high frequency oscillation periodic with one orbit revolution that are instead captured by the CR3BP that represents here the high fidelity reface dynamics. As expected the variation of semi-major axis predicted by the averaged approach is zero as visible in Figure 1a while the long-term variation of the other Keplerian elements is properly captured.

In order to better highlight the role of the Moon in the long term perturbation, we now focus our analysis on the variation of the orbital elements over one single orbit evolution and we consider different orientation of the Moon with respect to the orbit apse line. Ref shows the variation of inclination over one single revolution as predicted by the averaged approach and the CR3BP.

Table 1: Orbital elements of XMM-like spacecraft with respect to an Earth-centred reference frame with the x-y plane corresponding to the Earth-Moon plane.

| Keplerian | *a* [km] | *e* | *i* [deg] | Ω [deg] | ω [deg] |
|---|---|---|---|---|---|
| XMM spacecraft | 87736.3 | 0.82 | 57.2 | 108.9 | 270.9 |

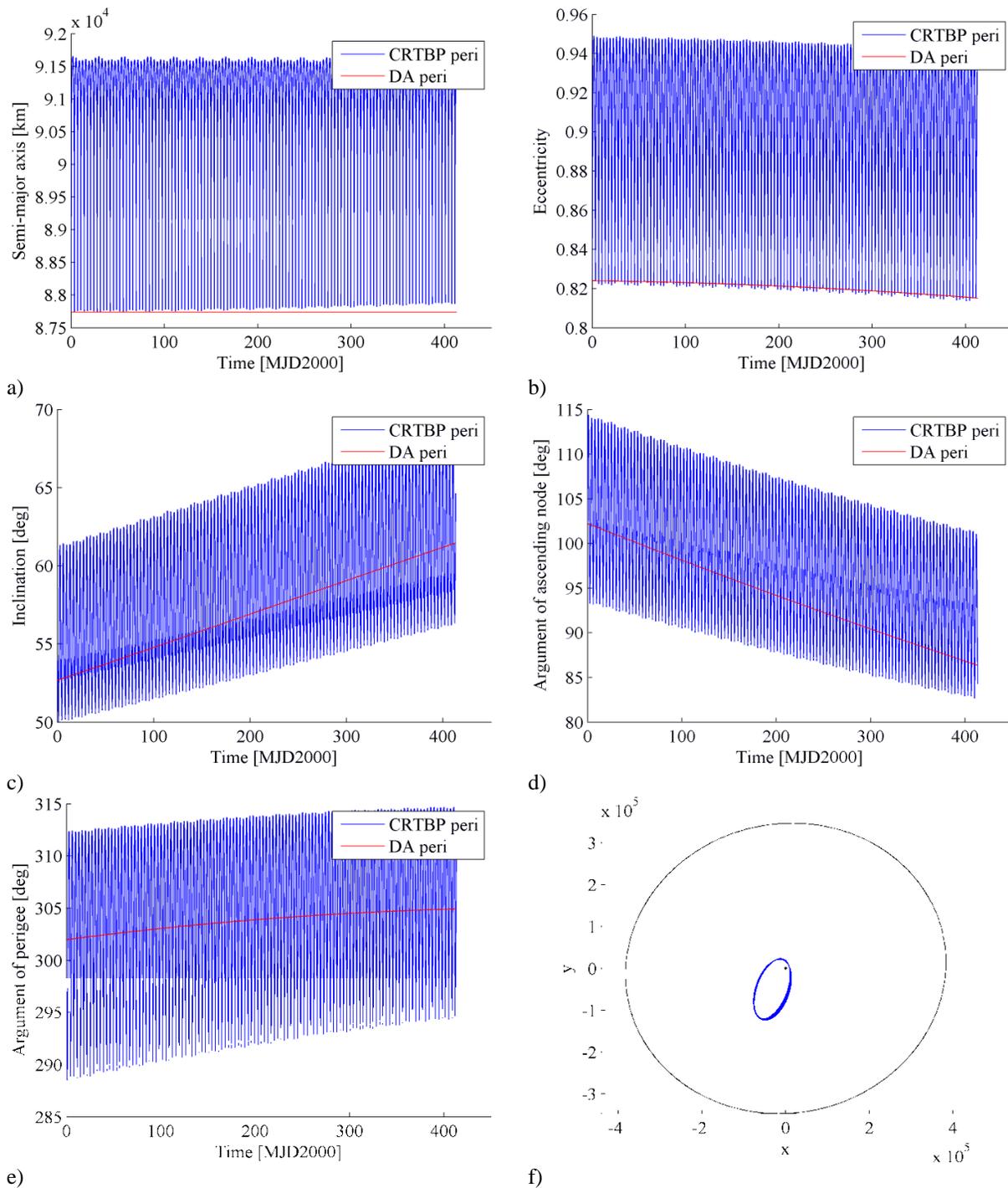

**Figure 1.** Long term evolution of the orbital elements of XMM spacecraft under the perturbing effect of the Moon using the semi-analytical double averaged method and the CR3BP.

A *Kick map* representation is thereafter referred as the variations of the osculating elements $\Delta a$, $\Delta e$, $\Delta i$, $\Delta \omega$ of a massless particle due to third body perturbations, as a function of the phasing between the orbit of the third body and the massless particle. Note that since the third body (the Moon in this test case) is assumed to move in a circular orbit, the axis used to account for both the third body's true anomaly and the longitude of the ascending node $\Omega$ of the massless particle (here, XMM spacecraft) can be defined arbitrarily. These two angles essentially define the relative angular phasing between the spacecraft and the Moon. Hence, the relative angular phasing between the massless particle and the perturbing body can be determined by resolving the difference between angular position of the massless particle at its periapsis, which is the result of a rotation defined by the Euler angles $\Omega$, $i$, $\omega$, and the angular position of the perturbing third body at the same epoch of the spacecraft's periapsis passage.

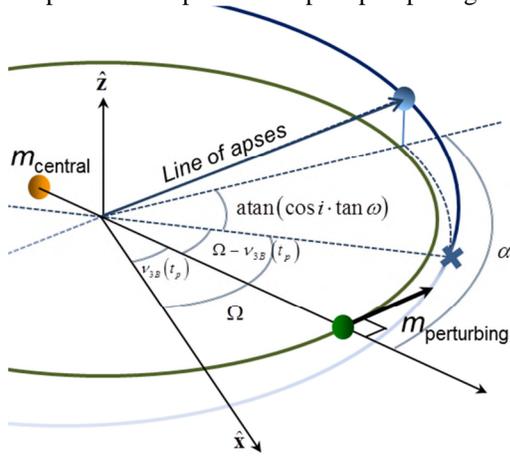

**Figure 2.** Schematic of the angular distance $\alpha$.

More particularly, the angle α is here defined as the angle between the line connecting the central body and the third body and the projection of the periapsis line of the massless particle onto the third body's orbital plane at the moment of the periapsis passage of the massless particle, measured from the system barycentre as shown in Figure 2. This can be computed as:

$$\alpha = \Omega - v_{3B}(t_p) + \arctan(\cos i \cdot \tan \omega) \qquad (9)$$

where $\Omega - v_{3B}(t_p)$ is the angular distance between the massless particle's ascending node and the position of the third body along its circular orbit at the moment of the massless particle's periapsis passage. Note that when computing the arctangent, one must choose the result at the same quadrant as for $\omega$.

Figure 3 shows the kick map representations for the changes in semi-major axis and eccentricity as a function of the angular phasing between the line of apses of XMM's orbit and the Moon when XMM is at its periapsis. Figure 3 shows then the changes of these two osculating elements, i.e. ($a$,$e$), as computed by the Keplerian Map approximation, but also the results of the numerical propagation of the same conditions with the CR3BP equations of motion. The initial conditions defined as a set of Keplerian elements are transformed into synodical coordinates and propagating both for half a period backwards and half a period forwards (red cross symbols). The final conditions are again transformed back to the Keplerian element space and the variations are computed by simply subtracting the initial value to the final value of the corresponding Keplerian element. By comparing these two sets of results, the accuracy of the Keplerian map can be assessed.

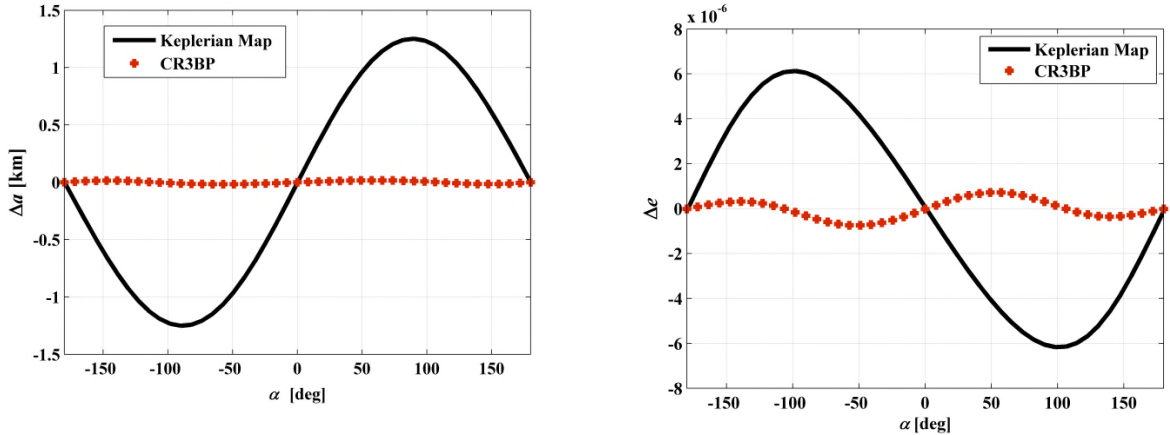

**Figure 3.** Kick maps for an XMM-like orbit.

It is clear that the perturbation is small, but also that the error of the Keplerian map approximation is notable. While further analysis needs to be carried out to determine the actual source of the problem, it is believed this is due to high $\mu$ of the Earth-Moon system, as well as the high eccentricity of XMM orbit, which, combined, render the linearly truncated binomial approximations of $r$ and $r_{3b}$ inaccurate. The following test case, however, presents a much more convenient scenario where Keplerian Map approximation renders an accurate description of the dynamics.

Asteroid 2011 BL45

We now focus our attention on asteroid 2011 BL45, which was discovered during its latest Earth approach, in 2011. The approach had a minimum encounter distance with the Earth of only 0.07 AU, according to NEODYS[†]. Prior to this encounter the asteroid had the following Keplerian elements:

---
[†] http://newton.dm.unipi.it/neodys

**Table 2: Orbital and Test parameters data.**

| Keplerian | $a$ [AU] | $e$ | $i$ [deg] | $\Omega$ [deg] | $\omega$ [deg] | $v$ [deg] | Epoch |
|---|---|---|---|---|---|---|---|
| 2011 BL45 | 1.0365 | 0.020 | 3.0523 | 135.15 | 157.19 | 68.285 | 22/06/2006 (00:00) |
| Synodic frame | $x$ | $y$ | $z$ | $v_x$ | $v_y$ | $v_z$ | |
| | 0 | 1.0277 | -0.0390 | 0.0387 | 0.0168- | -0.0376 | |
| Mass Parameter | $\mu$ | 3.0401x10$^{-6}$ | | | | | |

An orbit such as that of asteroid 2011 BL45 can now be used to test the accuracy of the Keplerian Map approximation, for a system in which the mass parameter $\mu$ is 3.0401x10$^{-6}$, which accounts for the mass fraction of the Earth and Moon in the Sun-Earth+Moon system.

Figure 4 then shows the kick map variations of the osculating elements $\Delta a$, $\Delta e$, $\Delta i$, $\Delta \omega$ along one complete orbital revolution of asteroid 2011 BL45. The orbital changes were computed at all possible orbital phasing conditions by varying $\Omega$-$v_{3B}(t_p)$, hence $\alpha$. The real periapsis passages during the last time this asteroid underwent a close encounter with the Earth were also plotted in the figure. It is clear that the accuracy of the Keplerian Map approximation is much better for this example, which represents an Earth-like orbit slightly perturbed by the Earth during a distant encounter ($>r_H$ where $r_H = (\mu/3)^{1/3}$).

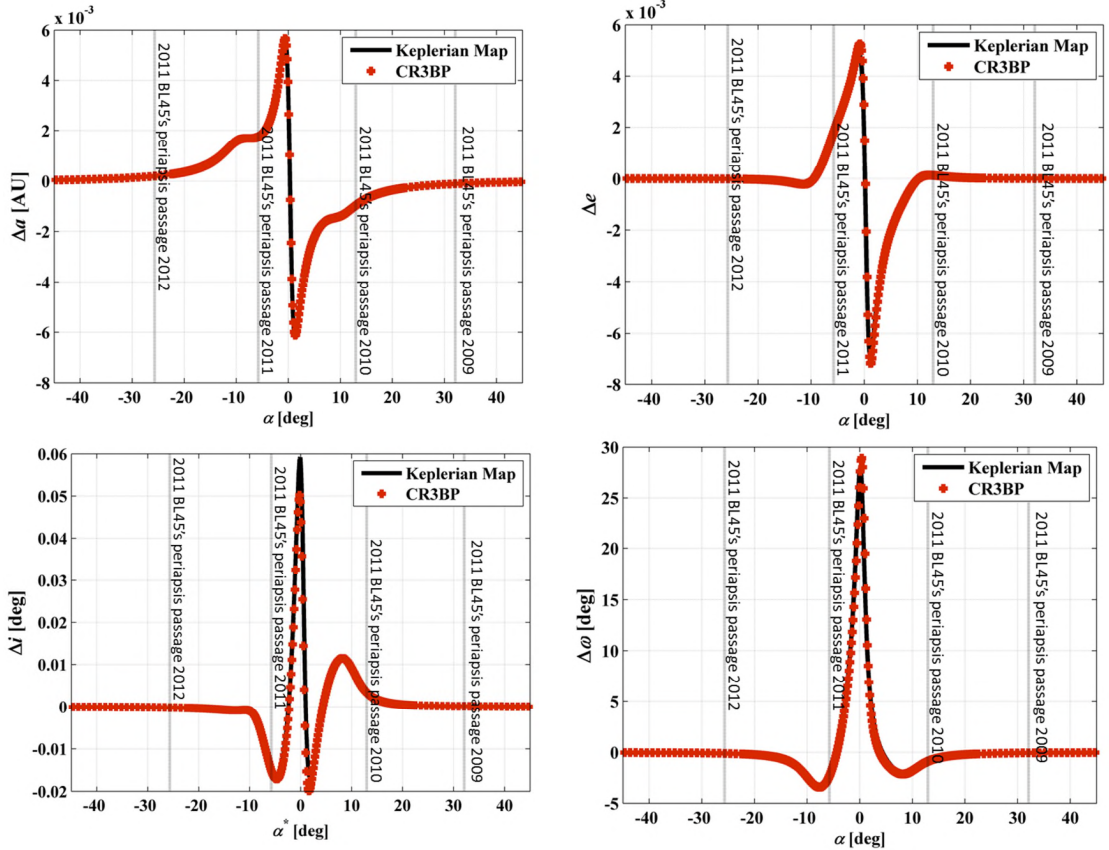

**Figure 4.** Kick maps for an asteroid '2011 BL45'-like orbit.

Figure 4 however represents only the differences between the final and initial conditions of a complete orbit. The accuracy of the Keplerian map approximation can also be assessed by looking into the time evolution of the osculating elements at some particularly interesting encounter geometries. For example, Figure 5 shows the evolution of the osculating semi-major axis as a function of the non-dimensional time for $\alpha$=-0.7°, which corresponds to the maximum $\Delta a$ perturbation in Figure 4. The initial conditions, as for the computation of the kick maps in Figure 3 and Figure 4, are taken at the periapsis and propagated forwards and backwards for half orbits, generating a complete orbit propagation. While the propagation of the CR3BP is done straightforwardly by numerically solving the ordinary differential equations, the propagation of the Keplerian map is done by solving the integrals described in Alessi and Sanchez [4] by means of a Picard method for a specified timespan.

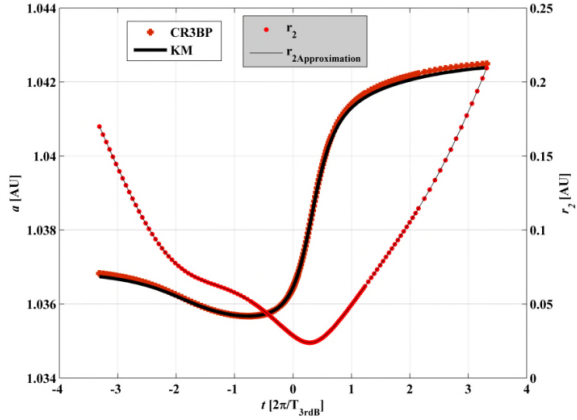

**Figure 5.** Time evolution of the osculating semimajor axis at α=-0.7° for an '2011 BL45'-like orbit.

Hence, Figure 5 again shows an accurate description of the Keplerian map approximation of the CR3BP, for this case. The figure also plots both the real distance to the secondary object, or third body, and the linearly truncated binomial approximation of it. Next, Figure 6 shows another interesting case, which corresponds to α=-0.1°, the encounter geometry that yields the maximum Δ$i$ in Figure 4. Note that this is the case where a maximum difference in the kick map representation can be observed between the Keplerian map results and the CR3BP propagation. Indeed, a difference can be observed between the osculating inclination as propagated by the CR3BP and that approximated by the Keplerian map. Yet the difference is still relatively small.

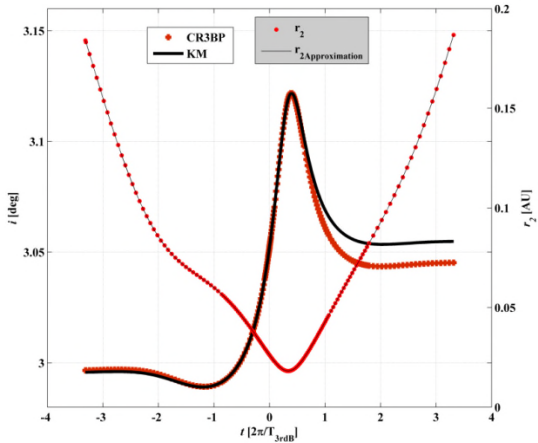

**Figure 6.** Time evolution of the osculating inclination at α=-0.1° for an '2011 BL45'-like orbit.

### Jupiter-Callisto System

To compare the results that can be obtained with the CR3BP, the Keplerian map and the Öpik's theory, we consider the Jupiter-Callisto system ($\mu \sim 5.6681 \times 10^{-5}$) and we show the change in ($a,e,i,\omega$) corresponding to different geometric configurations with respect to Callisto.

First, we notice that one of the main assumptions (stated earlier) required for Öpik's theory to be valid is that the Tisserand parameter $\mathcal{T}$ associated with the orbit must be less than 3. The simulations performed show that, at least for the system considered, $\mathcal{T}$ must be at least 2.9 in order to obtain an acceptable description of the consequences due to a close encounter, no matter on the distance of the approach. It turns out that if $\mathcal{T}$ is low and the distance is high, that is, grazing the Hill's sphere or even a little above, the theory can be still applied, while if $\mathcal{T}$ tends to 3 and the distance is very low it does not. This was expected, because the whole formulation is based on the hyperbolic dynamics.

In Figure 7 to Figure 9, we show three cases, corresponding to $\mathcal{T} \approx 2.18$ and $MOID \approx 0.0055$, $\mathcal{T} \approx 2.18$ and $MOID \approx 0.0266$ and $\mathcal{T} \approx 1.68$ and $MOID \approx 0.0027$ non-dimensional units, respectively. Recall that the radius of the Hill's sphere at Callisto is about 0.0266 non-dimensional units and the radius of the moon is about 0.00127 non-dimensional units. We notice that, at the configuration, i.e., α, corresponding to the closest approach, the Keplerian map and Öpik's theory give the same results. On the one hand, this is an evidence that the Keplerian map can work well also inside the sphere of influence. On the other hand, Öpik's formulation is much faster and should be preferable in these situations. Moreover, the main discrepancies of both methodologies with respect to the propagation of the CR3BP equations of motion are detected in the variation in semi-major axis. For the Öpik's theory, the variation in ω is reliable only at the closest approach, i.e. if the encounter with the moon is distant, then the variation obtained diverges substantially from the real one.

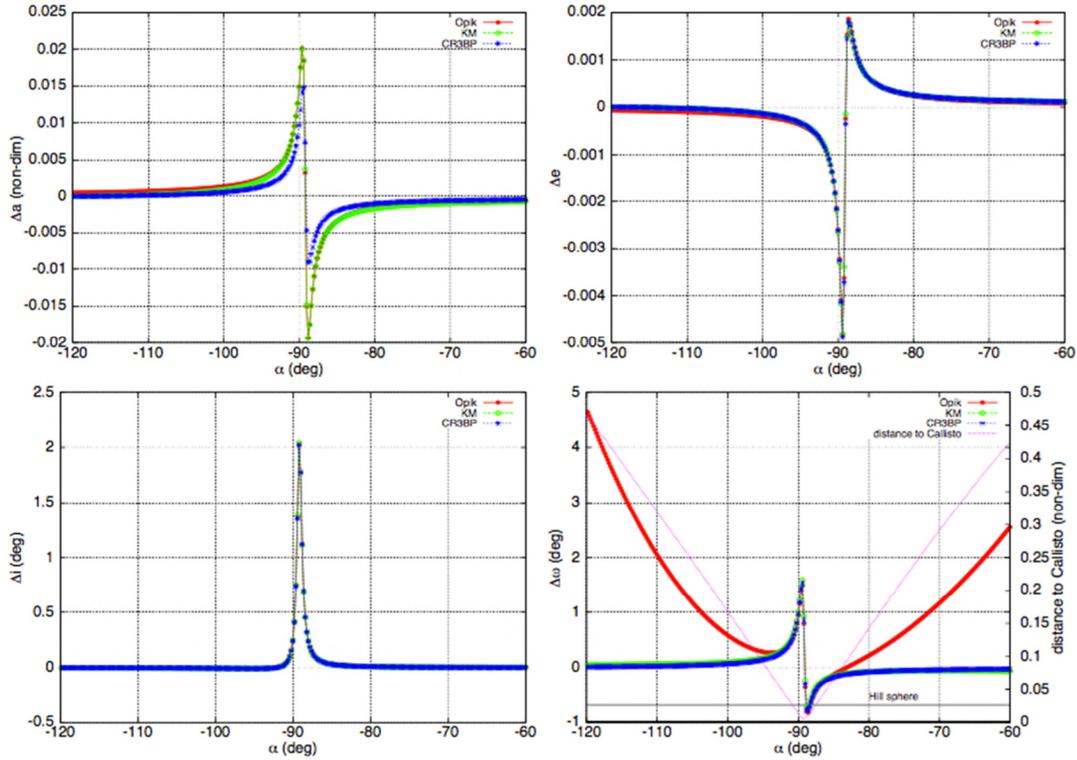

**Figure 7.** Change in ($a,e,i,\omega$) induced by Callisto according to the CR3BP, the Keplerian map and Öpik's theory, according to different relative configurations between the spacecraft and the moon, for an initial condition such that $\mathcal{T} \approx 2.18$ and $MOID \approx 0.0055$ in non-dimensional units.

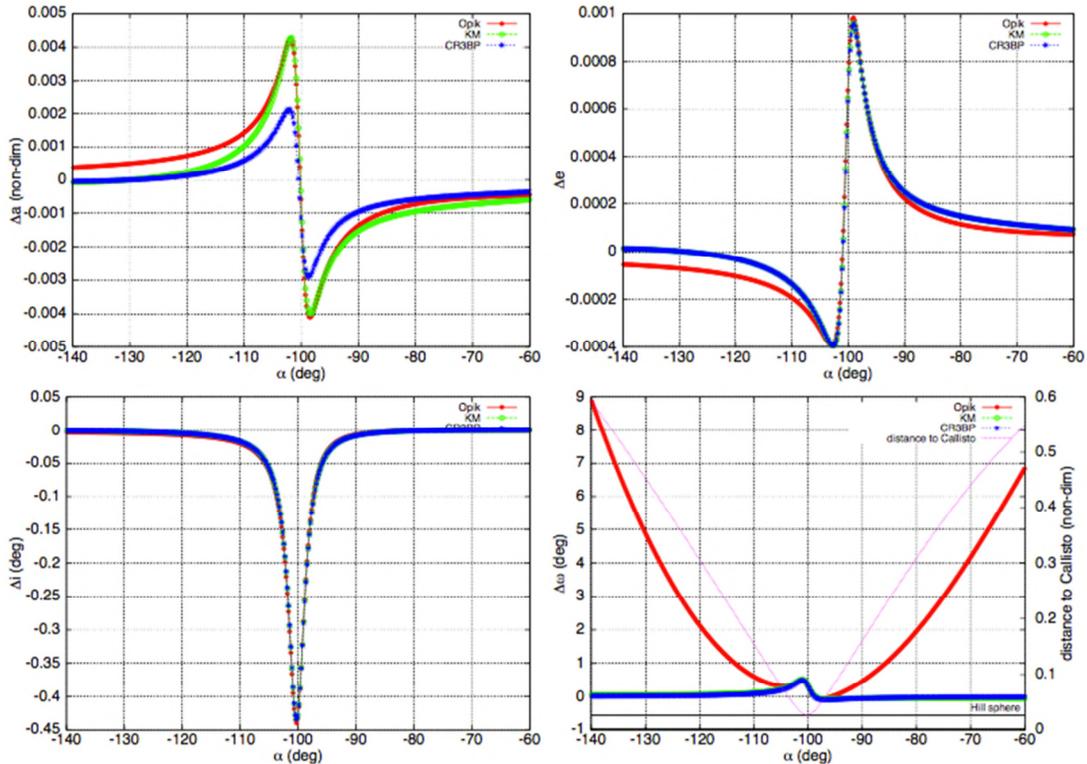

**Figure 8.** Change in ($a,e,i,\omega$) induced by Callisto according to the CR3BP, the Keplerian map and Öpik's theory, according to different relative configurations between the spacecraft and the moon, for an initial condition such that $\mathcal{T} \approx 2.18$ and $MOID \approx 0.0266$ in non-dimensional units.

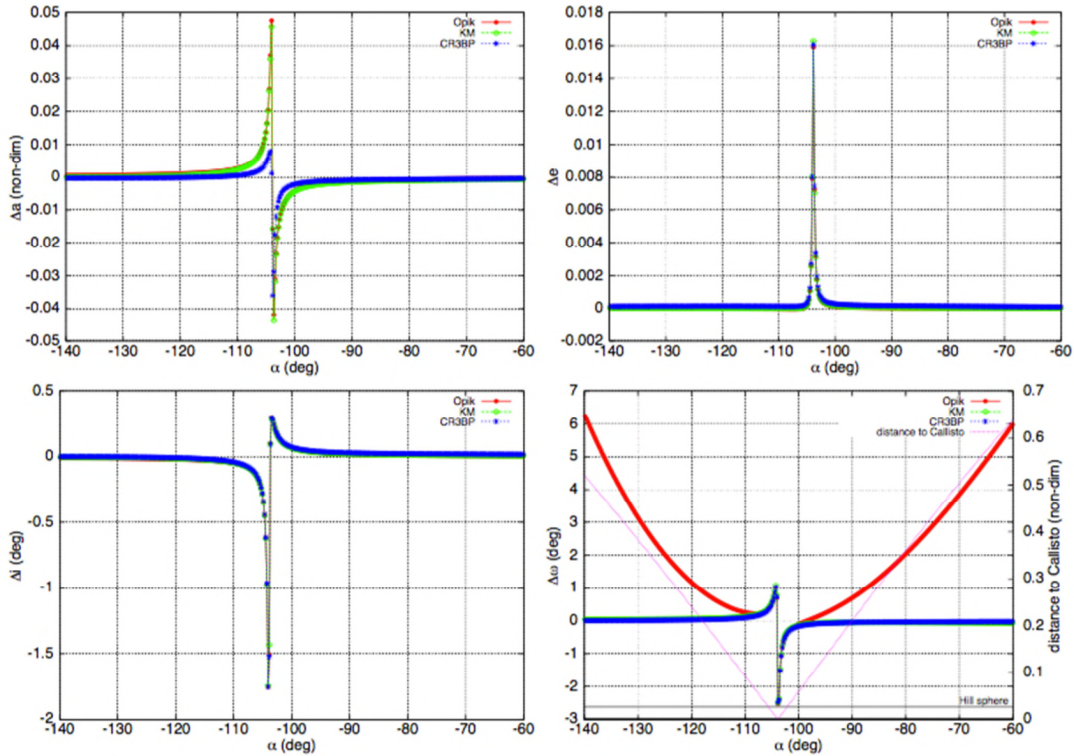

**Figure 9.** Change in ($a,e,i,\omega$) induced by Callisto according to the CR3BP, the Keplerian map and Öpik's theory, according to different relative configurations between the spacecraft and the moon, for an initial condition such that $\mathcal{T} \approx 1.68$ and $MOID \approx 0.0027$ in non-dimensional units.

## CONCLUSIONS

This paper has begun to explore the accuracy of three different methods to estimate the effects of the gravitational perturbation of a third body. These three methods were: the averaged dynamics applied to the third body perturbation, Öpik's close encounter theory and the Keplerian mapping approach. These three methods have been classically applied to three completely different regimes of motions. The third body approach, when the perturbing body always remains very far from the massless particle trajectory, Öpik's theory for extremely close encounters with the third body, well within the sphere of influence of the latter, and finally Keplerian mapping approach for trajectories that undergo distant encounters, at distances of several Hill's radius.

However, the results presented here indicate some overlap between the different methods, for example Keplerian map can also be used for encounters occurring within the Hill sphere, as long as the perturbation does not change drastically the original Keplerian motion. Öpik's theory also provides relatively accurate estimates for encounters occurring far from the third body, as long as the relative velocity is high. Clearly, however, the methods have limitations that occur when one or more of the approximations entailed in the method cease to provide a good approximation.

This work is an ongoing effort to provide a detailed knowledge of the regimes of applicability of each of these perturbative methods.